\documentstyle[aps,amssymb]{revtex}

\input{tcilatex}

\begin{document}
\title{Dynamical properties of the conformally coupled flat FRW model}
\author{M. A. Castagnino}
\address{Instituto de Astronom\'{\i }a y F\'{\i }sica del Espacio}
\author{Casilla de correo 67, sucursal 28}
\author{1428 , Buenos Aires, Argentina.}
\author{H. Giacomini}
\address{Laboratoire de Math\'{e}matiques et Physique Th\'{e}orique- UPRES A 6083}
\author{Facult\'{e} des Sciences et Techniques}
\author{Universit\'{e} de Tours-Parc de Grandmont}
\author{37200 Tours, France}
\author{and FCBF-Area de F\'{\i }sica, UNR}
\author{Suipacha 531, 2000 Rosario, Argentina.}
\author{L. Lara}
\address{Depto. de F\'{\i}sica, FCEIA, UNR}
\author{Avda Pellegrini 250, 2000 Rosario, Argentina.}
\author{lplara@arnet.com.ar}
\maketitle

\begin{abstract}
In this paper we study the dynamical behaviour of a simple cosmological
model defined by a spatially flat Robertson-Walker geometry, conformally
coupled with a massive scalar field. We determine a Lyapunov-like function
for the non-linear evolution equations. From this function we prove that all
the stationary solutions are unstable. We also show that all initial
conditions, different from the stationary points, originate an expanding
universe in the asymptotic regime, with a scale parameter $a(t)$ that goes
to infinity and the scalar field $\phi (t)$ that goes to zero in an
oscillatory way . We also find two asymptotic solutions, valid for
sufficiently large values of time. These solutions correspond to a radiation
dominated phase and to a matter dominated phase, respectively.
\end{abstract}

\section{Introduction}

In this work we briefly analyze some new properties of an oversimplified
cosmological model: a spatially flat Robertson-Walker geometry, conformally
coupled with a massive scalar field, which has been studied in many papers 
\cite{Futa}. The universe is well described by spatially flat ($k=0)$
cosmological models, which are also compulsory if we use inflationary
theories, and an inflation field is usually included in these models.
Moreover, the conformal coupling is necessary if we want to satisfy the
equivalence principle \cite{cacho}. Hence, the studied oversimplified
structure is necessarily contained in the universe, albeit complemented by
much more detailed features. Since this simplified stru

cture by itself implies several of the most important cosmological
properties we believe that the results of this letter are of interest.

These properties are:

a.- It expands to infinity for {\it all} non-stationary initial conditions.

b.- It contains a Lyapunov-like \footnote{%
A Lyapunov function is an evergrowing function of $t$ that vanishes at a
critical point and it is different from zero in a neighborhood of this
point. Our function is not a proper Lyapunov function because it is not of
definite sign.}function $F$ with a positive time derivative and therefore is
essentially time asymmetric.

c.- The final phase of the universe evolution is either radiation or matter
dominated.

d.- The scalar field $\phi (t)$ goes to zero in an oscillatory way in the
asymptotic regime.

\section{The Model}

Let us consider a flat Robertson-Walker geometry with scale factor $a(t)$
conformally coupled with a neutral massive scalar field $\phi (t)$ of mass $%
m.$ The hamiltonian reads \cite{Model}: 
\begin{equation}
{\cal H}=\frac{1}{2}\left( \stackrel{.}{\phi }^{2}-\stackrel{.}{a}%
^{2}\right) +\frac{m^{2}}{2}a^{2}\phi ^{2}=\frac{1}{m^{2}}k  \label{2.0}
\end{equation}
where 
\begin{equation}
k{\cal =}\text{ }\frac{1}{2}(\stackrel{.}{x}^{2}-\stackrel{.}{y}^{2})+\frac{1%
}{2}x^{2}y^{2}  \label{2.1}
\end{equation}
$x=m\phi ,$ $y=ma,$ $p_{x}=$ $\stackrel{.}{x},$ $p_{y}=\stackrel{.}{\,y}$
and the dots indicate derivative with respect to the conformal time $t,$
defined by the relation $dt_{p}$ $=$ $a(t)$ $dt$ , where $t_{p}$ is the
commoving or proper time. The field equations read:$\qquad $%
\begin{equation}
\stackrel{..}{x}\,=-\text{ }xy^{2},\qquad \stackrel{..}{y}\,=yx^{2}
\label{2.2}
\end{equation}
and the Einstein constraint is ${\cal H}=0.$ This condition expresses the
Hamiltonian constraint of general relativity. The fixed points are located
at: 
\[
x\text{ arbitrary,}\qquad y=p_{x}=p_{y}=0
\]
\[
y\text{ arbitrary,}\qquad x=p_{x}=p_{y}=0
\]
These fixed points are stationary solutions of the system.

\section{Main dynamical properties of the model}

Let us define the function: 
\begin{equation}
F(y,\stackrel{.}{y})=y\stackrel{.}{y}  \label{3.1}
\end{equation}

Taking the time derivative of this function along an orbit of the system and
using the second equation (\ref{2.2}) we obtain: 
\begin{equation}
\frac{dF}{dt}=\,\stackrel{.}{y}^{2}+\,x^{2}y^{2}\geqq \,0  \label{3.2}
\end{equation}
Hence, function $F$ is a monotonous growing function of $t.$ Then, we have 
\begin{equation}
\lim_{t\rightarrow +\infty }\text{ }F(y(t),\stackrel{.}{\,y}(t))=+\infty 
\label{3.5}
\end{equation}
for an arbitrary non-stationary initial condition. The case $%
\lim_{t\rightarrow +\infty }$ $F(y(t),\stackrel{.}{\,y}(t))=\,C$, where $C$
is a constant, is excluded since for great values of $t$ we have $y(t)\cong
\,\pm \sqrt{2Ct+C_{0}}$ and the second equation (\ref{2.2}) can not be
satisfied for real values of $x$. Here, $C_{0}$ is an integration constant
that appears when we solve the differential equation $y(t)\stackrel{.}{y}%
(t))\cong C.$ We conclude that $\lim_{t\rightarrow +\infty }$ $y(t)=\pm
\infty $, i.e. all the orbits in phase space $(x,y,p_{x},p_{y})$ go to
infinity. It is clear that eq. (\ref{3.2}) forbids the existence of periodic
solutions. If a periodic solution $x(t)\,,\,y(t)$ of period $T$ exists,
integrating both sides of eq. (\ref{3.2}) along this solution we obtain zero
in the left-hand side and a positive number in the right-hand side and we
have a contradiction. All initial conditions different from the fixed points
yield an universe evolution that end in and expanding phase.

Moreover, from this fact we see that all the fixed points are unstable. We
also conclude that the system has no chaotic behaviour because all the
orbits different of the fixed points are not bounded (see \cite{BB}).

Let us study now the asymptotic behaviour of the scalar field. For great
values of $t$ the function $y(t)$ goes to $\pm \infty $ in a monotonous way,
and the product $y\stackrel{.}{y}$ is positive. Then, the function $G(y,%
\stackrel{.}{y})=\,\frac{\stackrel{.}{y}}{y}$ is well defined and positive
in the asymptotic regime. Taking the time derivative of $G$ along an orbit
of the system and using the second eq. (\ref{2.2}) we obtain

\begin{equation}
\text{ }\stackrel{.}{G}\,=\,\frac{\stackrel{..}{y}}{y}\,\,-\,\frac{\stackrel{%
.}{y}^{2}}{y^{2}}=\,\frac{y^{2}x^{2}-\,\stackrel{.}{y}^{2}}{y^{2}}  \label{7}
\end{equation}

The constraint ${\cal H=}$ $0$ gives

\begin{equation}
\stackrel{.}{y}^2=\,\stackrel{.}{\,x}^2+\,x^2y^2  \label{8per}
\end{equation}

Replacing this expression of $\stackrel{.}{y}^{2}$in eq. (\ref{7}) we obtain

\begin{equation}
\stackrel{.}{G}\,=-\frac{\stackrel{.}{x}^{2}}{y^{2}}\leq \,0  \label{9per}
\end{equation}

Since the function $G$ is positive and always decreasing we have

\begin{equation}
\lim_{t\rightarrow +\infty }\text{ }G\,=\lim_{t\rightarrow +\infty }\text{ }%
\frac{\stackrel{.}{y}}{y}=k  \label{10per}
\end{equation}

where $k$ is a non-negative constant. Since function $G$ tends to a constant
value in a monotonous way, its first derivative must tends to zero when $%
t\rightarrow +\infty $ , i.e.

\begin{equation}
\lim_{t\rightarrow +\infty }\text{ }\stackrel{.}{G}\,=\lim_{t\rightarrow
+\infty }-\frac{\stackrel{.}{x}^{2}}{y^{2}}=0  \label{11per}
\end{equation}

Equation (\ref{8per}) can be written as follows

\begin{equation}
\frac{\stackrel{.}{y}^{2}}{y^{2}}=\frac{\stackrel{.}{x}^{2}}{y^{2}}+x^{2}
\label{12usan}
\end{equation}

Taking the limit when $t\rightarrow +\infty $ in both sides of this equality
and using eqs. (\ref{10per}) and (\ref{11per}), we obtain

\begin{equation}
\lim_{t\rightarrow +\infty }\text{ }x^{2}\,=k^{2}  \label{13cavallo}
\end{equation}

We will now show that the constant $k$ is zero. From eq. (\ref{13cavallo})
we have that $\lim_{t\rightarrow +\infty }$ $x\,=\pm k$. Let us consider a
first case where the limit is positive. For sufficiently great values of $t$%
, $x$ must approach the value $+k$ in a monotonous way. In fact, small
oscillatory behaviour with decreasing amplitude around the value $+k$ is
impossible, since such type of behaviour implies a sequence of changes of
sign of the second derivative of $x$. From the first equation (\ref{2.2}) we
see that it is impossible because the function $x$ does not change sign.

But, if the function $x$ approaches the value $+k$ in a monotonous way, its
first and second derivatives must go to zero when $t\rightarrow +\infty $.
Since $\lim_{t\rightarrow +\infty }$ $y^{2}\,=+\infty $, we see from the
first eq. (\ref{2.2}) that $\stackrel{..}{x}$ can not go to zero. The
analysis of the case where $x$ approaches the value $-k$ is analogous. Then,
we conclude that the constant $k$ must be zero, i.e. 
\begin{equation}
\lim_{t\rightarrow +\infty }\text{ }x\,=0  \label{14vanrrel}
\end{equation}

We have proved that $x$ can not approach the value $\pm k$ in a monotonous
way. The analysis is independent of the value of $k$. Then, we conclude that 
$x$ has an oscillatory behaviour with a decreasing amplitude when $%
t\rightarrow +\infty $. This oscillatory behaviour is not possible for $%
k\neq 0$ but it becomes possible in the case $k=0$ because $x(t)$ can change
sign.

To summarise, we have proved that the scalar field $\phi =\frac xm$ goes to
zero in a oscillatory way, the factor scale $a^2$ goes to infinity in a
monotonous way, and $\frac{\stackrel{.}{a}}a$ goes to zero when $%
t\rightarrow +\infty $ (eq.\ref{10per} with $k=0$).

\section{Asymptotic solutions}

For $t>>1$ we can obtain explicit asymptotic solutions. Instead of system (%
\ref{2.2}) we will consider the system: 
\begin{equation}
\stackrel{..}{x\,}=-xy^{2},\qquad {\cal H=}\text{ }0  \label{4.1}
\end{equation}
because deriving the second equation with respect to $t$ and using the first
one we obtain $\stackrel{..}{y}$ $=yx^{2}.$ System (\ref{4.1}) automatically
satisfies the constraint ${\cal H}=0$, while the system (\ref{2.2}) has
solutions that do not satisfy this constraint.

Inspired by exhaustive numerical calculations we have been able to determine
two different asymptotic solutions for $t>>1$, namely: 
\begin{equation}
x(t)\sim \frac{1}{t}\sin \left( \frac{1}{2}t^{2}\right) ,\qquad y(t)\sim t
\label{4.2}
\end{equation}
and 
\begin{equation}
x(t)\sim \frac{2}{t}\sin \left( \frac{1}{3}t^{3}\right) ,\qquad y(t)\sim
t^{2}  \label{4.3}
\end{equation}
Introducing these expressions in system (\ref{4.1}) and taking $t>>1$, it
can easily be verified that (\ref{4.2}) and (\ref{4.3}) are asymptotic
solutions. Expressions (\ref{4.2}) and (\ref{4.3}) agree with the general
results of section III.

Expressing these asymptotic solutions in terms of commoving time $t_p$ we
conclude that:

a.- The final phase of solution (\ref{4.2}) is radiation dominated (since $%
y(t_{P})\sim $ $t_{P}^{\frac{1}{2}})$.

b.- The final phase of solution (\ref{4.3}) is matter dominated (since $%
y(t_{P})\sim $ $t_{P}^{\frac{2}{3}})$.

We have proved that (\ref{4.2}) and (\ref{4.3}) are asymptotic solutions but
this does not imply that they are the unique solutions for $t\gg 1.$ For a
large number of initial conditions chosen at random, numerical computations
show that the associated asymptotic solution is given by (\ref{4.3}), i.e. a
matter dominated evolution.

We have performed our analysis in the conformal time $t.$ In general, to
translate the results to the commoving time $t_{p}$ is not an easy matter
(in some cases it is practically impossible) when the scale factor $a(t)$ is
not monotonous. In our case, for sufficiently large values of $t$, $a(t)$ is
monotonous and the asymptotic behaviour in the variable $t$ can be easily
translated to the commoving time $t_{p}$. If we choose physical initial
conditions in such a way that $a(t=0)\stackrel{.}{a}(t=0)>0$, it is easy to
show that $a(t)$ is a monotonous function of $t$. Then, for such initial
conditions all the results obtained in terms of $t$ can be easily translated
in terms of commoving time $t_{p}$, for arbitrary values of $t$. For more
general cosmological models where the scale factor does not has a monotonous
behaviour, the dynamical analysis in terms of conformal time $t$ is not
justified. In fact, it is clearly wrong from the point of view of dynamical
system theory.

\section{Conclusions}

In spite of the simplified nature of our cosmological model, its dynamical
evolution is controlled by highly non-linear equations. Nevertheless, we
have been able to obtain, in a rigorous way, the most relevant dynamical
properties of the model: the instability of all the stationary solutions,
the expansive nature of the evolution, the fact that the scalar field $\phi
(t)$ goes to zero in oscillatory way and explicit expressions for the
asymptotic solutions, giving the two possible phases of the universe.

Under the time inversion $t\rightarrow -t$ the hamiltonian (\ref{2.1}), the
field equations (\ref{2.2}) and the constraint ${\cal H}=0$ remain
invariant, so the system is trivially time-symmetric. Nevertheless, if the
motion begins at a finite point with a finite value of $F$ (and this would
be the case for the real universe that began very small and with high
temperature in a quantum phase), the motion always goes to infinity. Then,
it is time asymmetric since its initial and final states are different. Of
course, the inverse motion is also a solution of the evolution equations.
But the motion towards infinity is usually expanding and with creation of
matter-field energy, and therefore corresponds to what we see in the
observable universe.

\end{document}